\documentclass[cmbright]{staauth}

\usepackage{moreverb}

\usepackage[dvips,colorlinks,bookmarksopen,bookmarksnumbered,citecolor=red,urlcolor=red]{hyperref}
\usepackage{graphicx}
\begin{document}

\title{A Mutual Information Approach to Calculating Nonlinearity}

\author{Reginald Smith}

\address{%
\affilnum{a}PO Box 10051, Rochester, NY 14610}

\corremail{rsmith@bouchet-franklin.org}

\received{00 July 2015}
\accepted{00 July 2015}

\begin{abstract}
A new method to measure nonlinear dependence between two variables is described using mutual information to analyze the separate linear and nonlinear components of dependence. This technique, which gives an exact value for the proportion of linear dependence, is then compared with another common test for linearity, the Brock, Dechert and Scheinkman (BDS) test.
\end{abstract}


\maketitle

\section{Introduction}

The foundations of statistics, and most problems analyzing real data, grapple with the problem of accurately characterizing the dependence between two or more random variables (r.v.). This problem, over a hundred years old, was studied by luminaries of mathematics such as Gauss and Laplace but was put on a firm foundation by Galton's first expositions on correlation \cite{galtoncorr,galtonheight,galtonhist,galtonhist2}. The modern product-moment correlation coefficient was later designed by Karl Pearson and has been used in countless analyses. Correlation, which measures linear dependence, is only one of many measures including others such as Kendall 's $\tau$ \cite{kendall}, Spearman's ranked correlation coefficient $\rho_S$ \cite{spearman}, Blomqvist's $q$ \cite{blomqvist}, and information theoretic metrics such as mutual information $I$ \cite{shannon} and the normalized informational correlation coefficients $r$ of Linfoot \cite{linfoot} and $\delta$ of Joe \cite{joe}. 

All of these measurements have various benefits and limitations. The correlation coefficient $\rho$ is easy to calculate and interpret for positive or negative correlation, however, it measures only linear dependence. Kendall's $\tau$ and the rank correlation coefficient $\rho_S$ are more appropriate when using ranked data. Mutual information has the virtue of being the measurement of dependence that measures all (linear and nonlinear) dependence between r.v. but can be difficult to calculate, is prone to underestimation bias (like most information theoretic measurements), and has no readily clear interpretation.

Given the plethora of measures, some ideal aspects for a numerical measurements of dependence were outlined by R\'{e}nyi \cite{renyi} in a list of seven ``natural'' attributes for measures of dependence. For a measurement of dependence between r.v. $X$ and $Y$ designated by $\delta(X,Y)$

\begin{enumerate}
  \item $\delta(X,Y)$ is defined for any values of $X$ and $Y$ where neither r.v. is constant for all values (Existence property)
\item $\delta(X,Y)=\delta(Y,X)$ (Symmetry property)
  \item $1 \geq \delta(X,Y) \geq 0$ (Unit interval property)
  \item $\delta(X,Y)=0$ iff $X$ and $Y$ are independent (Zero independence property)
\item $\delta(X,Y)=1$ where there is strict dependence between $X$ and $Y$ such as $Y=f(X)$ or $X=g(Y)$ (One dependence property)
\item If the functions $f(X)$ and $g(Y)$ in the previous property map the real axis onto itself in a one-to-one mapping, then $\delta(f(X),g(Y))=\delta(X,Y)$ (Mapping property)
\item If the joint distribution of $X$ and $Y$ is normal then $\delta(X,Y)=|\rho(X,Y)|$ where $\rho(X,Y)$ is the correlation coefficient of $X$ and $Y$. (Normal correlation property)
\end{enumerate}

The problem of measuring nonlinearity has recently taken a forefront in the measurements of dependence due to the increasing realization that nonlinear dependence and phenomena are often as important, and likely more common, than simple linear dependence. The last condition in R\'{e}nyi's list, the normal correlation property, is the only one to address nonlinearity in a measurement of dependence and there it does so only indirectly. The problems of measuring nonlinearity are many. Unlike with linear dependence, using parametric models to measure nonlinearity is fraught with difficulty and uncertainty given an infinite selection of joint distributions \cite{frechet}. Nonparametric models, however, can be hobbled by their lack of specificity; they can rule out linearity but not propose an alternate model. They also may require different tests for different types of nonlinear dependence.

The most comprehensive measure of dependence, mutual information, is the best for measuring the total dependence between two variables but its interpretation is not widely understood and its application in simple methods of data analysis is still relatively rare. In this paper, however, we will use mutual information to show that it can successfully not only distinguish linear and nonlinear dependence but can also be used to show the relative proportion of dependence due to linear dependence (or any other selected model). While this is not a new measure of dependence itself, it can help evaluate the linear or other different components of dependence between variables.

\section{Correlation and linear dependence}
The history of the idea of correlation is closely tied to the idea of dependence and can be variously traced to multiple mathematicians such as Gauss, Laplace, Bravais, and Adrain \cite{corrdepend}. However, these were conceptual or theoretical derivations with limited connection to empirical data or practice. The true development of correlation as a tool and technique for empirical analysis was made by Francis Galton over the course of over a decade \cite{galtonhist,galtonhist2}. These included early research from 1874 on a scatterplot of the head sizes vs. heights for 92 members of the Royal Society in his \emph{English Men of Science} as well as his famous derivation of the bivariate normal distribution for parent and child heights \cite{galtonheight}. This research culminated in his 1888 paper ``Co-relations and their measurement'' \cite{galtoncorr}. This paper formally introduced the coefficient of correlation and defined correlation simply defining that variables are co-related when: ``the variation of the one is accompanied on the average by more or less variation of the other, and in the same direction.''

While Galton did calculate the first correlation coefficients, the modern formulation of the product moment correlation coefficient was left to Karl Pearson to formulate in 1896 \cite{pearsoncorr}. The correlation coefficient, the measure of linear co-variation of variables, was widely used, sometimes erroneously. Indeed, part of the wide use of the linear correlation coefficient, and the discarding of aspects of data that did not neatly fit it, derived from the limitations of the time. Yule \cite{yule} stated as much 

\begin{quotation}
\emph{In point of fact, few statistics would seem worth the labour of calculating any characteristic more complex than the linear. In most cases, the deviation from the linear character, at all events near the middle of the table where frequencies are greatest, does not appear to be very serious even though well defined.} 
\end{quotation}

Some of the first work done to explain the limitations of the correlation coefficient as a measure of dependence, in this case a measure of independence when $\rho=0$, was done by H.L. Reitz \cite{reitz}. He showed how in the example of simple harmonic motion $y=\cos(\lambda x)$ the overall correlation coefficient is zero though there is dependency between the two variables. Additional examples, including some geometric ones, are given in \cite{corrdepend}.

The Pearson product moment correlation coefficient, $\rho$, between two r.v. is calculated as

\begin{equation}
\rho = \frac{Cov(X,Y)}{s_X s_Y}
\label{correq}
\end{equation}

In equation \ref{correq}, $s_X$ and $s_Y$ are the sample standard deviations of r.v. $X$ and $Y$ respectively and $Cov(X,Y)$ is the covariance between $X$ and $Y$ defined as

\begin{equation}
Cov(X,Y)=\frac{1}{N-1}\sum_{i=1}^N (X_i-\bar{X})(Y_i-\bar{Y})
\label{cov}
\end{equation}

where $\bar{X}$ and $\bar{Y}$ are the sample means of $X$ and $Y$.

While correlation is undoubtedly the most widely used and understood measure of statistical dependence, it has limitations which are often misconstrued or even glossed over. The correlation coefficient is only a measure of linear dependence between two r.v. Any nonlinear dependence is not reflected in the value of the correlation coefficient. Though $\rho$ cannot overestimate dependence, it can underestimate dependence by nonlinear contributions. This additional dependence can be more subtle and not always in the same direction as correlation. A famous illustration of the shortfalls of the correlation coefficient in measuring dependence amongst variables is Anscombe's Quartet \cite{anscombe} where four plots of various, non-random data points (linear dependence, quadratic dependence, straight line with outliers, and vertical line with outliers) are shown that have wildly different forms of dependence but the same correlation coefficient.

\section{Current methods of measuring nonlinear dependence}

Measuring nonlinear dependence has been a frequent, but often only incompletely achieved goal. Early suggestions such as in \cite{nonlinear1} suggest using alternative regressions such as quadratic or cubic to see if these produce a better fit than linear regression. This would imply nonlinear dependence. 

The accelerated study of nonlinear dynamics and chaos in the 1980s and 1990s introduced several new tests. These later tests analyzed the residuals of linear regressions in order to determine if systematic departure from independent, identically distributed (i.i.d.) behavior could be found. Often, these are based off of hypothesis testing where the null hypothesis is the linear model and a test statistic is produced to accept or reject the null hypothesis at a given level of significance. These include the test of surrogate data \cite{surrogate} and the Brock, Dechert and Scheinkman (BDS) test \cite{bds1,bds2}. An extensive list of tests in use up to about 1990  and in the 1990s is given in \cite{nonlinear2} and \cite{nonlinear3}. 

Currently, the most widely used  test for nonlinearity is likely the BDS test statistic \cite{bds1,bds2}. The BDS test is a non-parametric test where the null hypothesis of i.i.d. residuals is analyzed using the correlation integral, a technique borrowed from nonlinear dynamics. In short, the correlation integral calculated on a time series measures whether the norm between two lagged groups of consecutive points are absolutely different by less than a selected small value, $\eta$. The number of points from each group compared at each step is known as the embedding dimension, $m$. If the residuals are i.i.d. the correlation integral should indicate an exponential decay of the correlation between two points as the lag between them increases. A slower decay of correlation typically indicates some sort of nonlinear dependence that may invalidate the null hypothesis. The output of the subsequent calculations in the BDS method is a test statistic that is normally distributed $N(0,1)$. This test statistic can then be used to accept the null hypothesis of i.i.d. residuals in the time series' autocorrelation, indicating linearity, or reject the null hypothesis with a given significance, typically $p$<0.05.

BDS is a sensitive test, non-parametric, and works for a wide variety of nonlinear relationships \cite{bds1,bds2}. However, in order to calculate the correlation integral, $m$, and $\eta$ (usually a multiple of the data's standard deviation) have to be chosen and must necessarily be adapted to the data set. This can lead to values that are sometimes arbitrary. As a statistical test though, BDS does have the advantage of allowing a statistically significant threshold to reject the null hypothesis. 

However, recently new approaches have been proposed to move away from the dominance of hypothesis testing. The quandary of defining measures of dependence for arbitrary models of data where the measures of dependence follow a condensed set of properties, similar to those of R\'{e}nyi, was taken up in \cite{reimherr}. Here, the authors note the inadequacy of both simple linear measures of dependence as well as hypothesis tests with only binary (0 and 1) but not intermediate results. They introduce the possibilty of developing custom information based (though not necessarily mutual information) models of dependence for specific problems.

\section{Mutual information in measuring total dependence}

In publishing his landmark paper, \emph{A Mathematical Theory of Communication}, Claude Shannon introduced information theory, and a variety of new statistical metrics, to the world. Chief among these is entropy, and its variants of joint, conditional, and relative entropy, as well as mutual information. Mutual information was introduced in the paper under the context of the channel capacity and maximum rate of transmission in a noisy channel given the source entropy and the conditional entropy between source symbols. The capacity of a noisy channel is the maximum mutual information amongst the symbols in the source. The mutual information can be understood as the reduction of uncertainty between one or more r.v. by knowing another. In other words, it is a measure of total dependence. Mutual information $I(X,Y)$ is defined both according to marginal ($H(X)$, $H(Y)$) and joint entropies ($H(X,Y)$)

\begin{equation}
I(X,Y) = H(X) + H(Y) - H(X,Y)
\label{MIsimp}
\end{equation}

as well as based on marginal and joint probability density functions (p.d.f.'s)

\begin{equation}
I(X,Y) = \int_{-\infty}^{\infty} \int_{-\infty}^{\infty} f(X,Y) \log \frac{f(X,Y)}{f(X)f(Y)} dX dY
\end{equation}

What makes mutual information a comprehensive measure of dependence is that it is based on any dependence of the r.v. that deviates from random chance. If two r.v. are independent, their joint entropy equals the sum of their marginal entropies. When the joint entropy is less than the sum of the marginal entropies, it indicates some form of dependence. However, unlike correlation, mutual information is non-parametric and assumes no sort of underlying distribution or  mathematical form of dependence. Therefore it is superior in taking into account both linear and nonlinear dependencies between r.v.

An example similar to Anscombe's quartet can illustrate the comprehensive measurement of dependence that mutual information can provide. Table \ref{MIquartet} shows several plots of random variables which have both nonlinear and linear dependence (as well as noise) but the same mutual information

\begin{table*}
\caption{An illustration of four types of dependence: positive linear, negative linear, sinusoidal, and bell shaped, all which have a mutual information of 1.32 nats (1.90 bits)}
\label{MIquartet}
\centering
 \begin{tabular}{cc}
 	 \includegraphics[height=2in, width=2in]{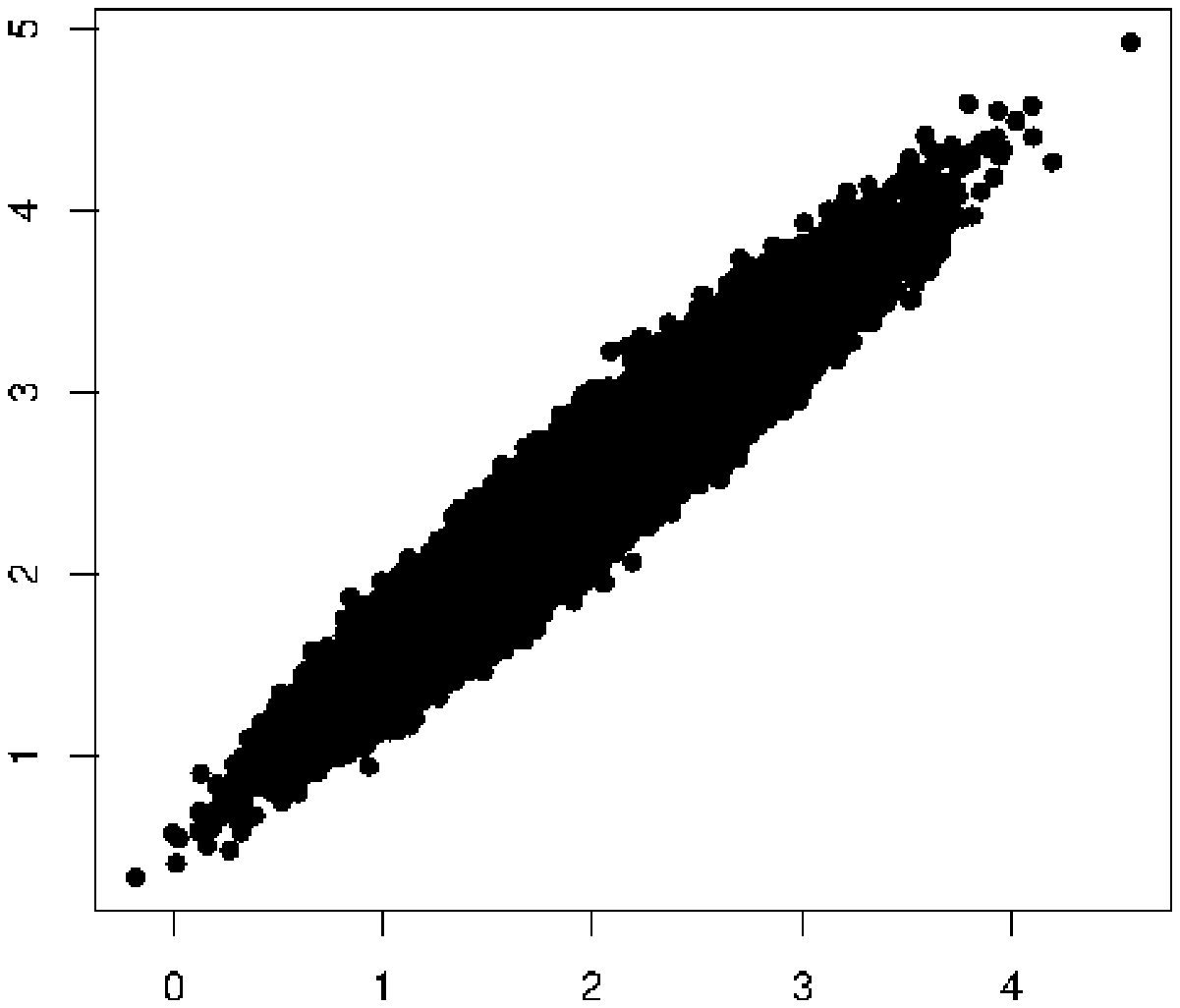}&
    \includegraphics[height=2in, width=2in]{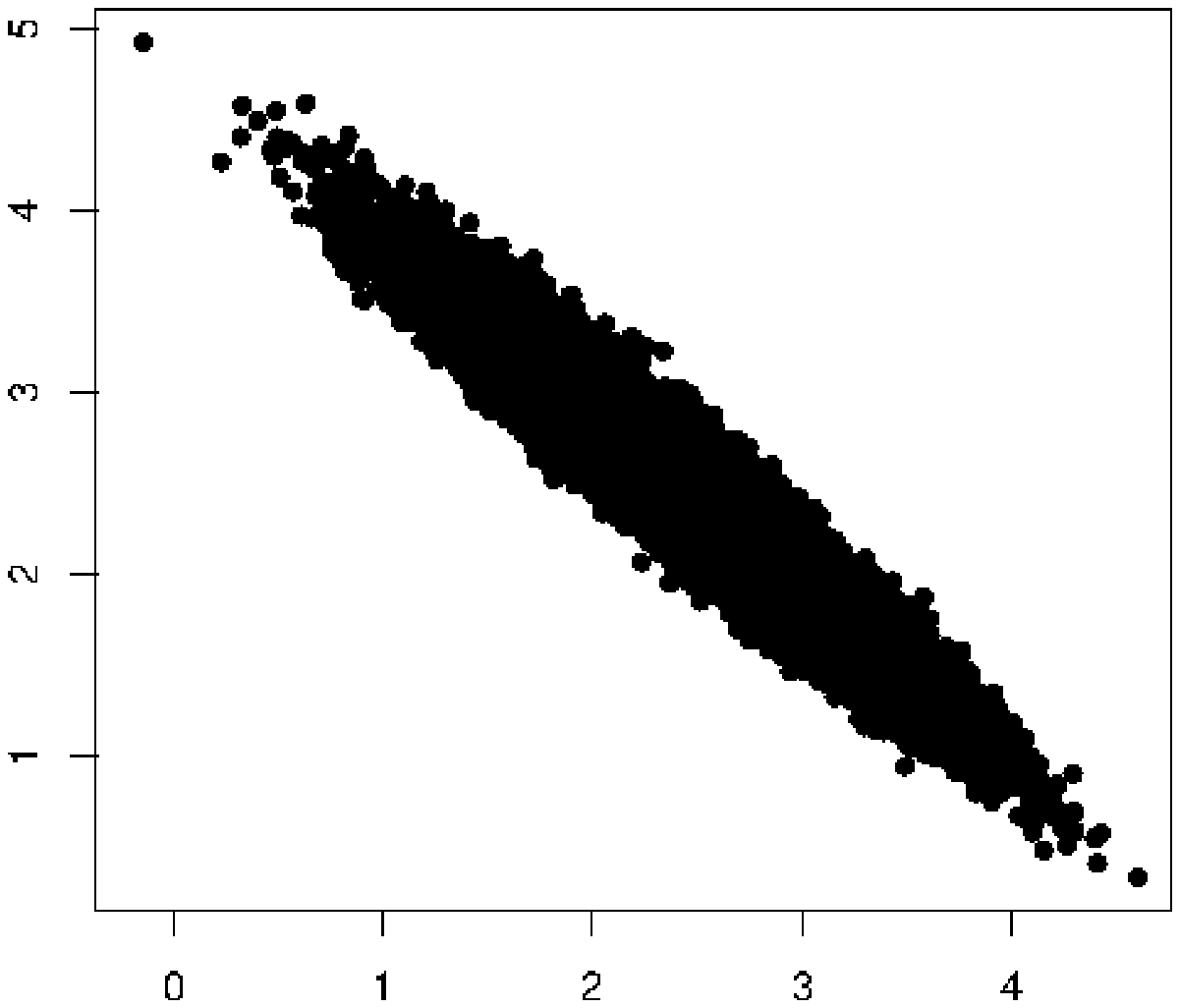}\\
\small{I =1.32 nats, $\rho$=0.96}&\small{I =1.32 nats, $\rho$=-0.96}\\
\includegraphics[height=2in, width=2in]{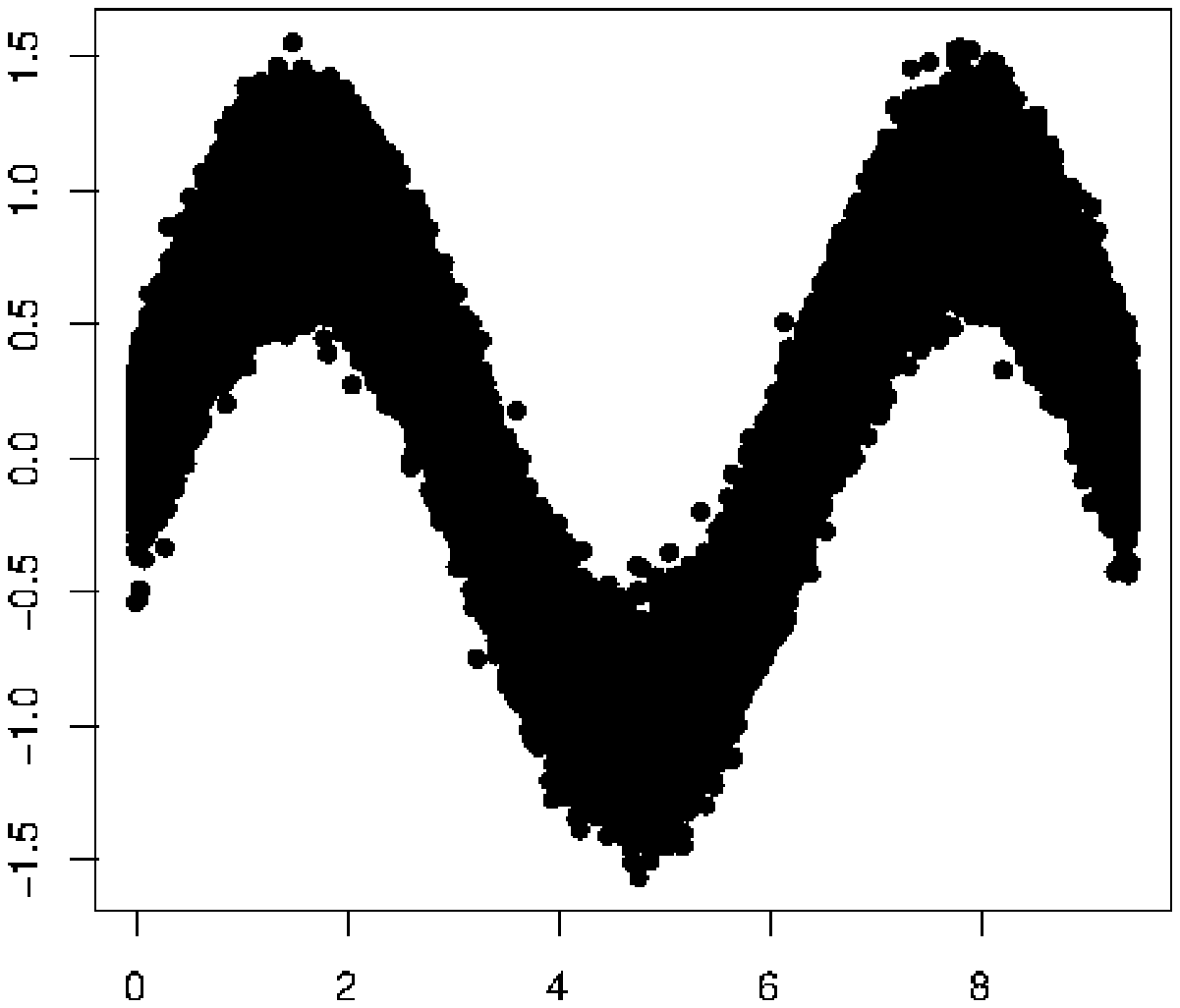}&
    \includegraphics[height=2in, width=2in]{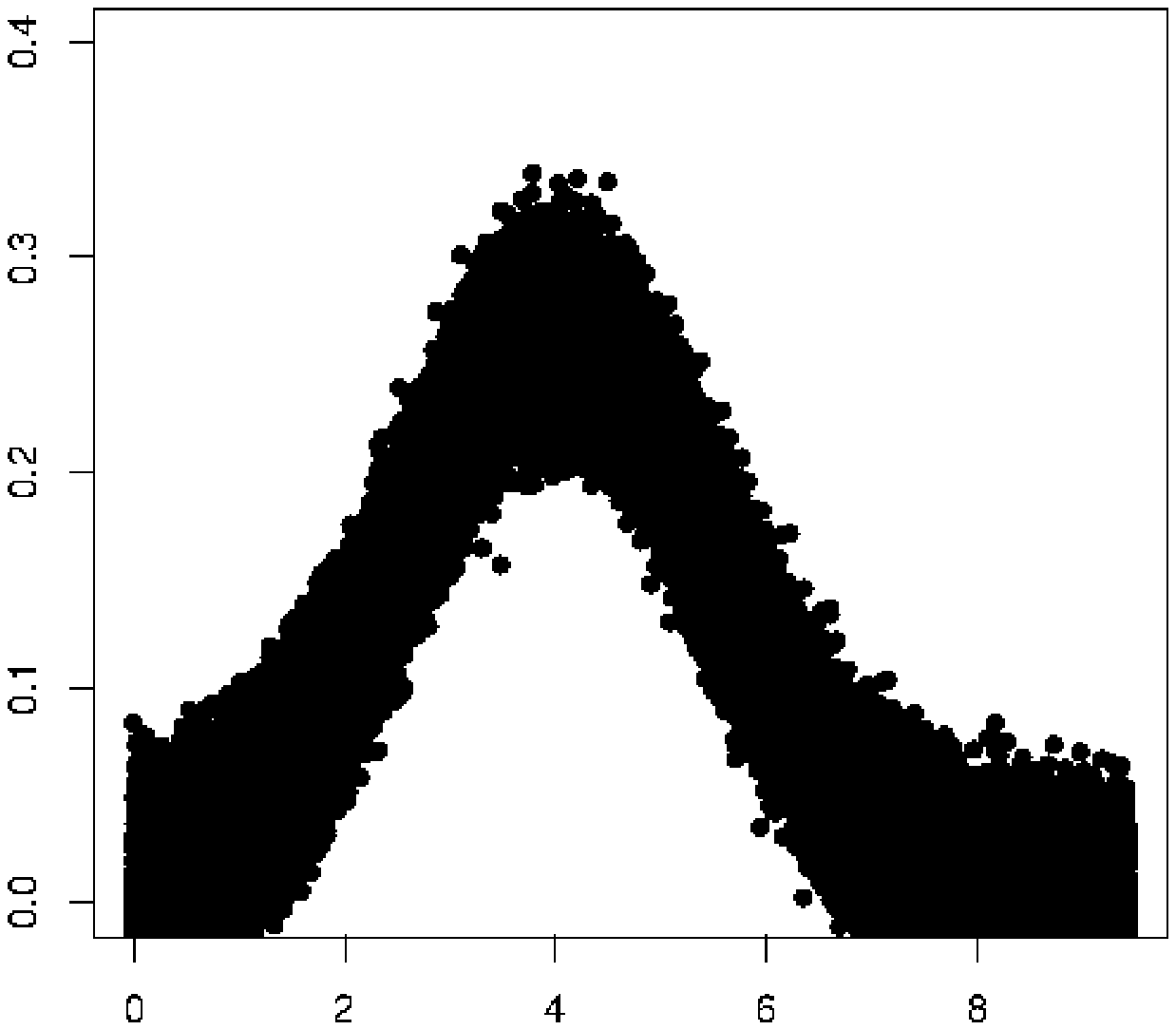}\\
\small{I =1.32 nats, $\rho$=0}&\small{I =1.32 nats, $\rho$=0.28}\\
 \end{tabular}
\end{table*}

Mutual information's importance was quickly recognized and it became a well-used measurement in coding theory, communications engineering, and even the physical and biological sciences. However, mutual information has several drawbacks in calculation and interpretation that have restricted its wider use. 

First, unlike correlation, which has an absolute value within the range of 0 to 1, mutual information's value is more open ended and can range from 0 for complete independence to infinity for a completely correlated and continuous pair of r.v.  The actual value of mutual information can vary with the values of the marginal entropies so there is no straightforward interpretation of a mutual information of 0.8 nats like there is with a correlation of $\pm 0.8$. While values of mutual information can be compared for different variable pairs in absolute terms, the maximum possible mutual information again varies with the marginal entropies of the r.v.

Second, mutual information, being based on probabilities, is difficult to calculate from empirical data. While various methods such as binning \cite{bin} or nearest neighbor measurements \cite{MImeasure2} are often used it is often not straightforward to take two data series and calculate the mutual information.  Third, the calculation of information theoretic entropies and mutual information suffers from well-known underestimation biases depending on the number of symbols and the size of the data set \cite{bias1,bias2,bias3}.

The investigation of biases in entropy (and thus mutual information) traces to the work of Miller and Madow \cite{bias1} who first investigated the problem. Their analyses, while correct, can sometimes underestimate the level of bias \cite{bias2,bias3}. The basic bias correction for entropy estimation by Miller and Madow is 

\begin{equation}
\hat{H} = H_{MLE} + \frac{m-1}{N}
\label{miller}
\end{equation}

In the equation above, $\hat{H}$ is the bias corrected entropy, $H_{MLE}$ is the maximum likelihood estimate based off of the empirical data, $N$ is the number of data points, and $m$ is the number of non-empty bins with data points.

Despite these issues, expressions relating mutual information and the correlation coefficient can be derived for a couple of restricted situations. Where the two marginals are normal distributions and their joint distribution is bivariate normal, the equivalence between mutual information and correlation is given by \cite{MIcorreq}

\begin{equation}
I = -\frac{1}{2} \log (1-\rho^2)
\end{equation}

The mutual information and correlation functions over a distance $d$ of an infinite, left-right symmetric binary symbol sequence can also be related as shown in \cite{wentian}. Defining the correlation function over length $d$ as $\Gamma (d)$, $P_0$ as the probability of a 0 in the sequence, and $P_1$ is a probability of a 1, under the conditions where $\frac{\Gamma (d)}{P_0P_1}$ is small and as $d \rightarrow \infty$, $\Gamma (d)$ decays to zero, we have the approximation

\begin{equation}
I(d) \approx \frac{1}{2}\big(\frac{\Gamma (d)}{P_0P_1}\big)^2
\end{equation}

There is no general relationship between $I(d)$ and $\Gamma (d)$ for more than two symbols.

Instead of equating correlation and mutual information under limited conditions, the analysis below should help complement these previous efforts by demonstrating the general correspondence between mutual information and correlation for any function or type of dependence. By demonstrating the proportion of dependence which is linear, it can both check the significance of linear regressions in the scheme of overall dependence as well as illuminate the nonlinear nature of dependence between variables.

\section{Calculating mutual information}

As previously stated, one of the key barriers to the wider use of mutual information to study dependence is the difficulty in calculating it from empirical data. The methods to calculate mutual information, as well as their respective pitfalls, are not widely known or studied--even in treatments on information theory. This section will serve as a brief introduction to the  most widely used methods and discussion of their relative merits. For more detailed analysis, the reader is directed to the references.

Like entropy, and other information measures, the most common and well-known method of calculating mutual information from empirical data is to use binning to create an approximate probability density distribution that can be used to calculate mutual information. The calculation proceeds either directly from the two variables by estimating the marginal and joint distributions or it proceeds indirectly by calculating the marginal and joint entropies and calculating mutual information using equation \ref{MIsimp}. While sample sizes can influence underestimation biases, the selection of the bin sizes is crucial in producing the most accurate result given the data. An overview of the effect of bin sizes on entropy estimation is given in \cite{bin}. In general, bin sizes should not necessarily be equal width but selected to contain about the same number of data points across each bin. This can be achieved a number of ways from sizing bins based off of the distribution assumed in the null hypothesis to creating a group of equal sized bins and merging adjacent ones until the equal frequency requirement is met.

However, there are several caveats to be recognized with binning when using it to estimate information theoretic measures such as entropy or mutual information. Two of these are discussed in \cite{MImeasure1} where it is demonstrated that if the ratio of data points, $N$, to non-zero bins, $m$ is 1 or less, the estimate will be biased even if corrections such as equation \ref{miller} are used. In addition, if both $N$ and $m$ are large, error bars for confidence intervals may be inaccurate and underestimated \cite{MImeasure1}. A final issue is that even if binning is done optimally, entropy and mutual information require calculating the logarithms of frequencies to estimate the logarithms of real probabilities. This can introduce error, especially where frequencies are very small. The second, and newer method, is an adaption of $k-$nearest neighbor statistics to estimate entropy. Introduced in \cite{MImeasure2} this technique holds the advantage of providing some correction for the errors introduced by sparsely populated bins.

\section{Estimating the linear component of mutual information}

Having discussed both linear correlation and mutual information, we can now discuss a possible means to combine the best features of each.  For such an analysis to be generally valid it must be non-parametric for both the marginals of the two variables as well as the joint distribution.

Using the correlation coefficient to estimate the linear component of mutual information will provide us with several key insights. First, it will allow us to clarify the full level of dependence above and beyond that shown by correlation and least-squares regression. Second, it can help provide an indirect evaluation of models assuming a linear causal dependence as the primary interaction between the two variables. Third, it can help integrate mutual information, and thus a greater knowledge of dependence, into analyses where it has previously been deemed an unfit descriptor.

Our technique in this paper will be a top-down approach to measuring the dependence of the variables. Given two r.v., $X$ and $Y$ we will define their marginal p.d.f.'s as $f(X)$ and $f(Y)$ respectively as well as a joint p.d.f. $f(X,Y)$. Their respective marginal entropies will be designated as $H(X)$ and $H(Y)$ and their joint entropies and mutual information by $H(X,Y)$ and $I(X,Y)$.

The ultimate goal is to measure the mutual information of the data as originally presented, remove the linear component of dependence as measured by the correlation coefficient, recalculate the mutual information on the new data set, and compare the mutual information of the data from both data sets.

\subsection{Step 1: Correlation coefficient calculation and least-squares regression}

The first step is not a radical departure from current methods of analysis. Namely, calculating the correlation coefficient $\rho(X,Y)$ and the equation for the least-squares regression. Let us describe the regression function in terms of predicting $Y$ given $X$. We will designate as $\hat{Y}$ the fitted values of $Y$ and the residuals for each value of $Y$ are designated as $z$

\begin{equation}
z_i=Y_i - \hat{Y_i} 
\end{equation}

\subsection{Step 2: Calculating the mutual information of the original variables}

Next, we calculate the mutual information $I(X,Y)$. In this treatment, the mutual information is calculated between the two variables using the equal-frequency binning discretization from the R package \emph{infotheo} \cite{R,infotheo} along with the mutual information calculation routine. The Miller-Madow bias correction was applied to the calculated values.

The value of $I(X,Y)$ will be used again in the last step for comparison.

\subsection{Step 3: Analysis of the residuals}

The residuals $z_i$, in a normal analysis, are investigated graphically and with statistical tests to determine if they have a mean of zero and whether their mean or variance is constant over the entire data set, etc. In our analysis, since we are not assuming that the relationship is linear, the residuals serve another purpose. Namely, the residuals represent the nonlinear dependence between $X$ and $Y$ and thus become the new focus of analysis in their own right. By understanding the the mutual information between the independent variable, $X$, and the residuals $z$ we can distinguish between linear and nonlinear components of dependence between $X$ and $Y$.

By definition, for a linear relationship the correlation between the residuals and the independent variable $X$, $p(X,z)=0$. It would seem that a logical next step would be to calculate $I(X,z)$, however, this mutual information is not directly comparable to $I(X,Y)$. In order to compare the mutual information between $X$ and $z$  and $X$ and $Y$ in an accurate way, we would want the full difference of the mutual information to be contained in the joint entropy between each pair of variables.

In other words, the mutual information comparison should be between two pairs of variables: $(X,Y)$ and $(X,Y')$ where $Y$ and $Y'$ both have the same marginal p.d.f. and thus the same marginal entropy so that $H(Y)=H(Y')$. This way, we can ensure that the difference in mutual information between $I(X,Y)$ and $I(X,Y')$  is only due to different joint entropies--one which has a linear component and one which does not. If the marginal entropies $H(Y)$ and $H(z)$ are different, it makes it unsure whether the mutual information difference between $I(X,Y)$ and $I(X,z)$ is due to the entropy of $z$ or the joint entropy $H(X,z)$. To calculate $Y'$, we must normalize the values of $z$ to create a distribution $Y'$ whose entropy matches $Y$.

This can be accomplished through a technique first known as the van der Waerden normal quantile transform \cite{quantile1,quantile2,quantile3,quantile4} and known in microarray data analysis as quantile normalization \cite{quantile5,quantile6}. Originally invented to map the cumulative distribution function (c.d.f.) of an arbitrary function onto the c.d.f. of the normal distribution for analysis, we use it here to map the c.d.f. of the residuals from linear regression onto the c.d.f. of the dependent variable. This allows us to ensure the mapped residual c.d.f., and thus probability density function, match those of the dependent variable and have the same entropy. This accomplishes the goal of the previous paragraph of reducing the differences in mutual information to differences between the joint entropies $H(X,Y)$ and $H(X,Y')$.

The quantile transform calculation of $Y'$ follows as

\begin{equation}
Y'=F_y^{-1}(G(z))
\label{quantilenorm}
\end{equation}

where $F_y$ is the c.d.f. of the original dependent variable $Y$, $F_y^{-1}$ is the quantile function of $Y$ and $G(z)$ is the c.d.f. of the residuals, $z$. By this method, $Y'$ shares the p.d.f. and entropy of $Y$ but is based on the values in the residuals. It can be shown that invariably, if $\rho(X,z)=0$ then $\rho(X,Y')$ is also equal to zero. Given equation \ref{quantilenorm} provides a direct dependence between $Y'$ and $z$, $\rho(Y',z)=1$ so it follows both have zero correlation with $X$.

\subsection{Step 4: Calculate mutual information of $X$ and $Y'$}

As a final step, we can calculate the mutual information of $X$ and $Y'$ which we will designate as $I'$. The relationship of $I'$ and $I$ can be succinctly described via the data processing inequality $I(X,Y) \geq I(j(X), k(Y))$ given $j(X)$ and $k(Y)$ are arbitrary functions of the original variables. Thus it is clear that $I \geq I'$.

The mutual information $I'$ can be understood as the mutual information of $X$ and $Y$ minus their linear dependence component. Thus, if linear dependence almost completely describes their relationship, $I' \approx 0$. If linear dependence is small or nearly non-existent, $I' \approx I$. The overall proportion of the linear dependence to all dependence between $X$ and $Y$ will be measured by $\Lambda$ where

\begin{equation}
\Lambda = 1 - \frac{I'}{I}
\end{equation}

It can  be quickly seen that $\Lambda$ has several attractive properties. First, the range of $\Lambda$ is restricted to [0,1]. Second, $\Lambda=0$ means there is no linear dependence between variables, but it does not mean $I=0$. If $I>0$ it indicates all dependence is completely nonlinear. Third, if $\Lambda=1$, the relationship is entirely linear and can be completely described by the linear correlation coefficient.

\section{Example: Anscombe's Quartet}

One quick and readily useful example is using the techniques above to measure the proportion of dependence which is linear in each of the Anscombe quartet plots. Given they all have the same value of the correlation coefficient, $\Lambda$ should be able to distinguish between them despite this similarity. Indeed this is the case. For Anscombe's Quartet recreated in four plots, each of which is made of $N$=10,000 points and measuring mutual information using $m$=50 equal frequency bins, the bias corrected mutual information was calculated and $\Lambda$ was derived. Results are shown below in Table \ref{Anscombe2}.

\begin{table*}
\caption{Anscombe's Quartets of $\rho$=0.693 shown with the respective $\Lambda$ for each.}
\label{Anscombe2}
\centering
 \begin{tabular}{cc}
 	 \includegraphics[height=2in, width=2in]{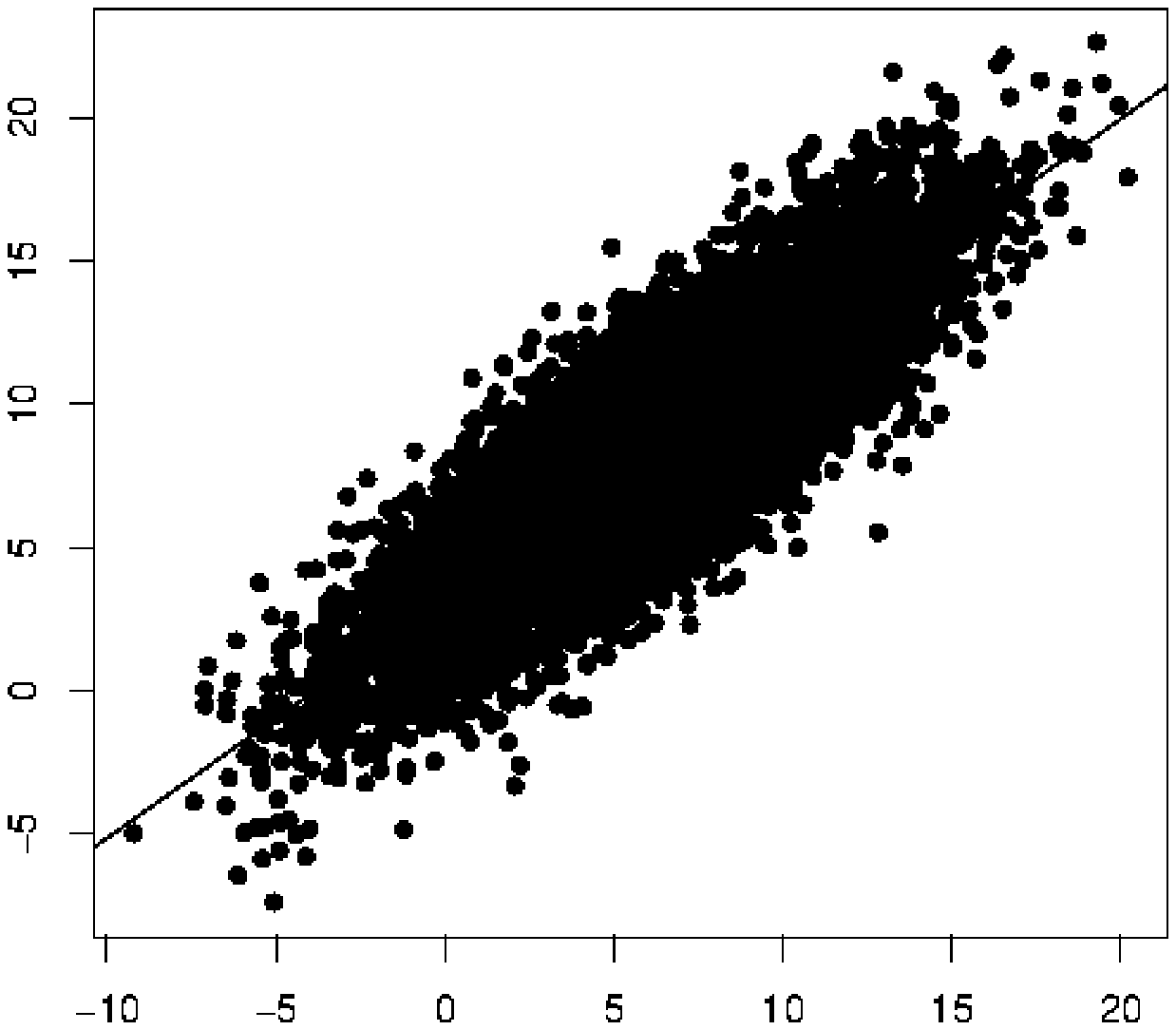}&
    \includegraphics[height=2in, width=2in]{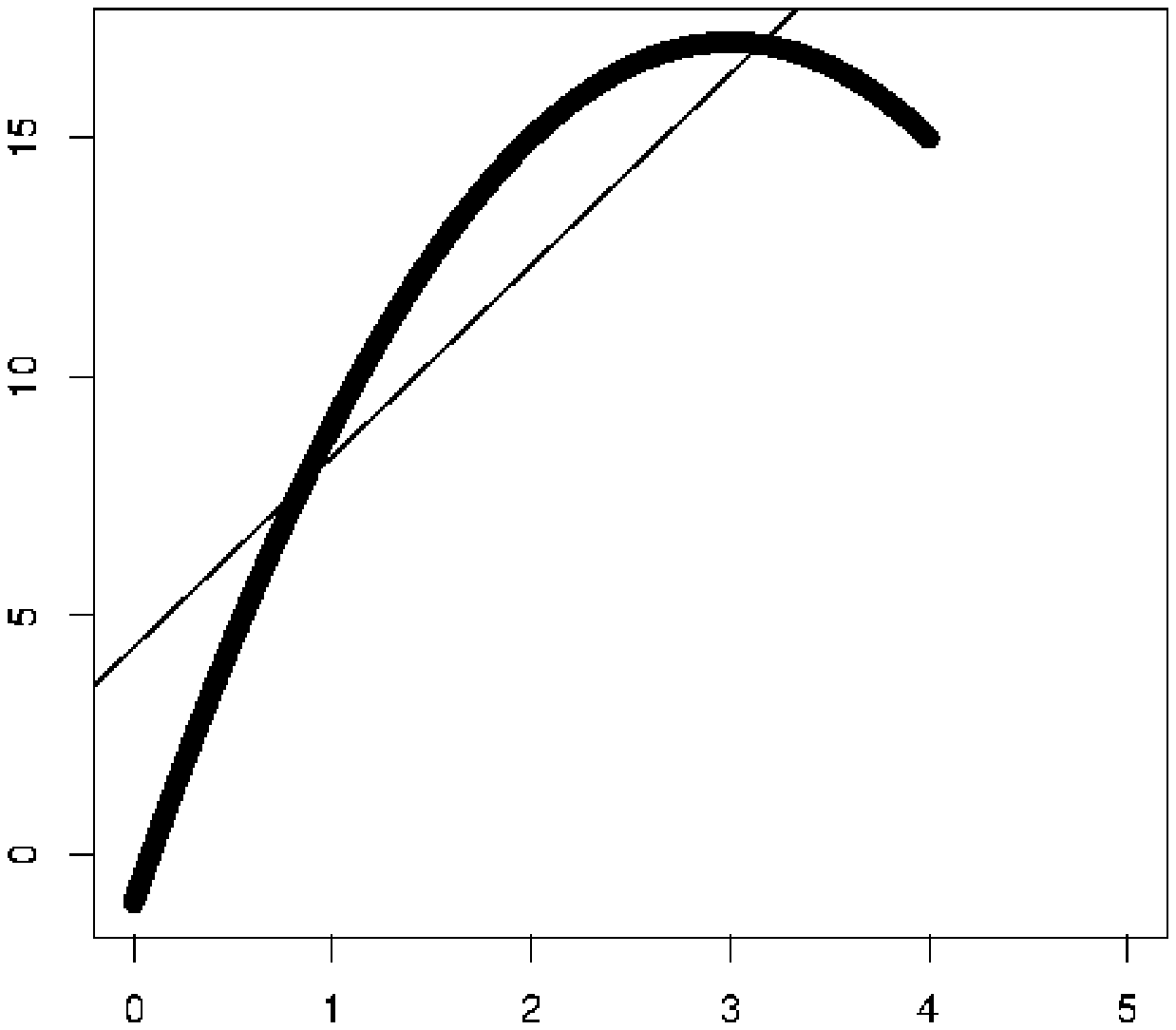}\\
\small{$\Lambda$=0.98}&\small{$\Lambda$=0.1}\\
\includegraphics[height=2in, width=2in]{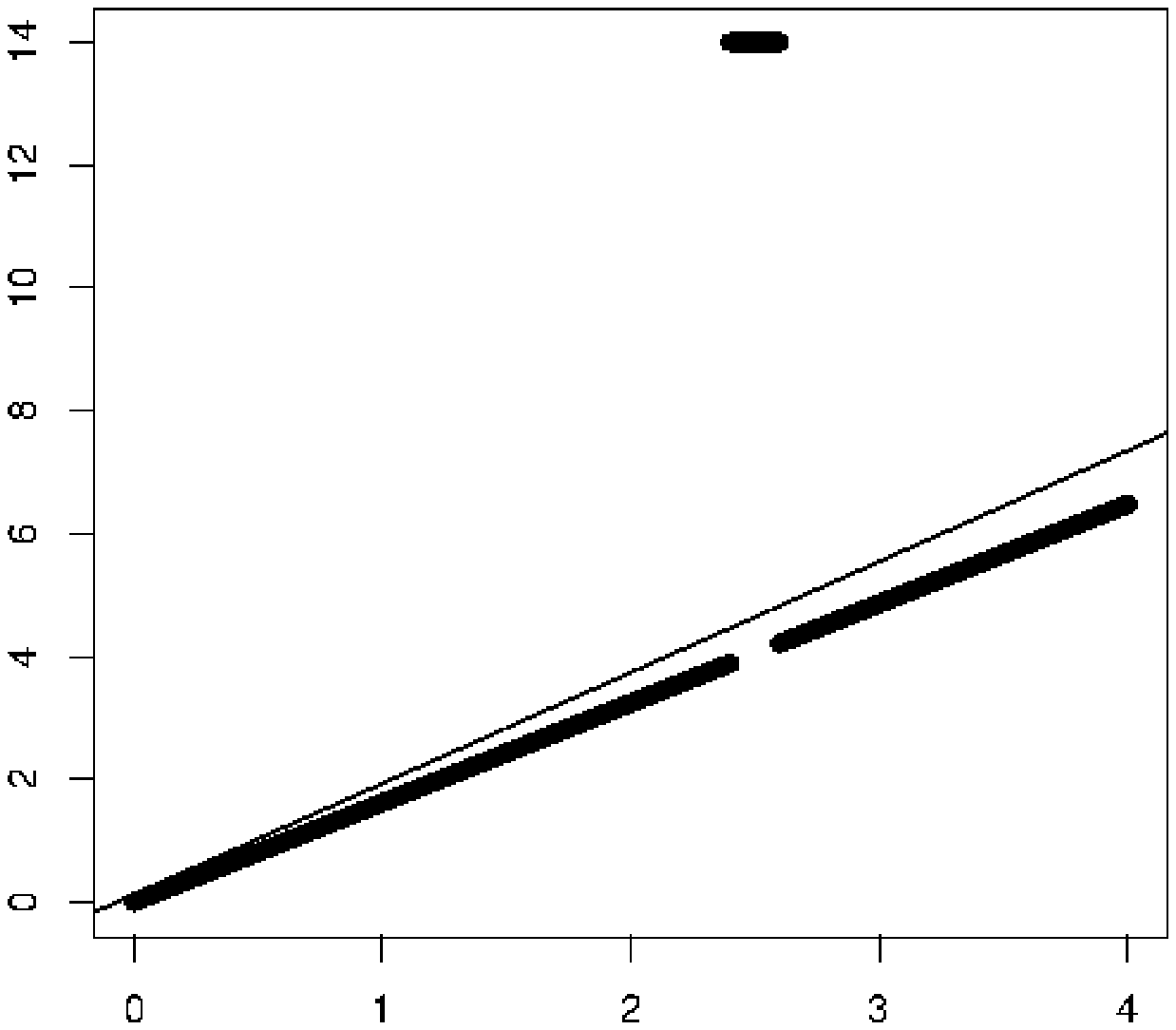}&
    \includegraphics[height=2in, width=2in]{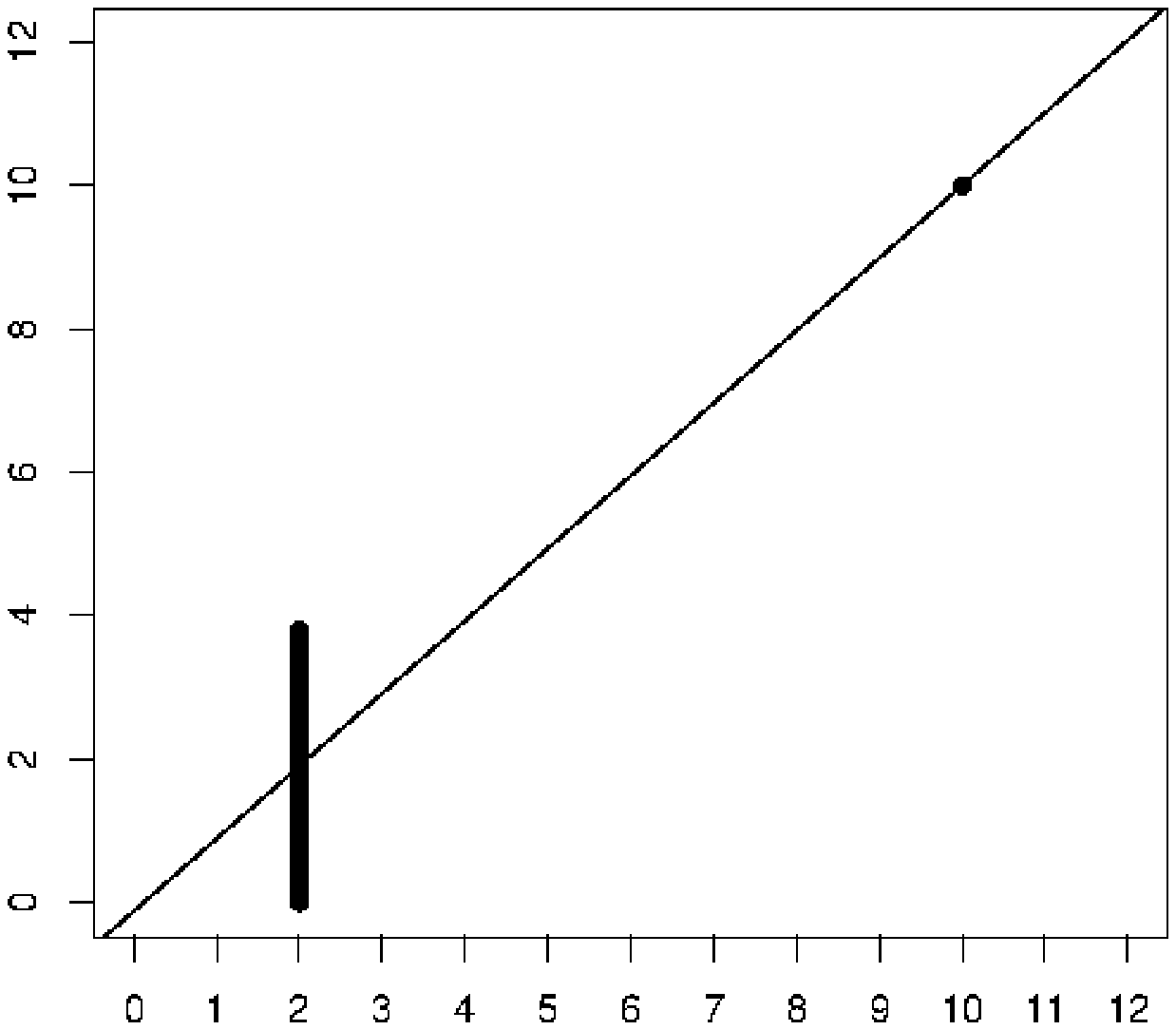}\\

\small{$\Lambda$=0.05}&\small{$\Lambda$=0.05}\\
 \end{tabular}
\end{table*}

\subsection{Using other fitted models}

The general, non-parametric nature of the calculation of $\Lambda$ lends itself to using any fitted model, not just linear regression, to create the residuals to use for subsequent calculation. In this case, $\Lambda$ measures not the proportion of total dependence that is linear, but the proportion of dependence explained by the fitted model. However, while a model may explain most of dependence due to fit, $\Lambda$ does not say whether the model is appropriate.

\section{Comparison with other tests for nonlinearity}

$\Lambda$, while indicating contribution to dependence, has no clear cutoff where a model can be considered, linear, nonlinear, or even a mix of both. The BDS test, and others, on the other hand are hypothesis tests which clearly give a significance level for the null hypothesis of linearity. These two methods of linearity can be used in a complementary fashion, however, with the BDS test giving an indication of nonlinearity and $\Lambda$ dictating the size and significance of linearity and nonlinearity in the data.

In an attempt to understand the typical cutoff for linearity in data, $\Lambda$ can compared to the BDS statistic for two functions. Two r.v. with $N=10,000$ points were created with the independent variable being a $N(0,1)$ distributed r.v., with a noise term represented by $\xi$,  and the dependent variable having quadratic and cubic dependence for the respective data sets. For data sets 1 and 2, the variables $X$ and $Y$ were respectively:

\begin{eqnarray}
X_i = \xi_i \nonumber \\
Y_i = 3X_i + aX_i^2 + \xi_i \nonumber\\
\end{eqnarray}

\begin{eqnarray}
X_i = \xi_i \nonumber \\
Y_i = 3X_i + aX_i^3 + \xi_i \nonumber\\
\end{eqnarray}

Above $a$ is a coefficient that is increased, starting at 0, to model increasing nonlinearity in the dependence of $Y$ and $\xi$ is the $N(0,1)$ noise term. The BDS test was applied to the residuals of linear regression between the two variables.

In both cases, $a$ was increased in small increments until the BDS test $p$ value at embedding dimension $m=2$ and $\eta=0.5$ became $p<0.05$. At this threshold value of $a$, $\Lambda$ was then calculated. In both data sets, the BDS test was shown to be quite sensitive to nonlinearity with $\Lambda = 0.95$ being the approximate value where the null hypothesis of i.i.d. residuals was rejected for $p<0.05$. Therefore, it is likely that linear models, as indicated by current tests of nonlinearity, imply a high value of $\Lambda$ near 1.0. However, as $\Lambda$ indicates, when $\Lambda$ < 0.95 the null hypothesis of linearity is often rejected in cases where there is a large component of linearity in addition to nonlinear effects. Therefore, $\Lambda$ could be useful in understanding the overall structure of dependence between variables and elucidating their various components, both linear and due to other functions.

This linear component of dependence can, however, complicate the analysis of dependence in certain conditions. For example, take an exponential dependence. Using the following relationship:

\begin{eqnarray}
X_i = \xi_i \nonumber \\
Y_i = e^{0.3X_i} + \xi \nonumber\\
\end{eqnarray}

$\Lambda=0.43$ for linear regression. Looking at the residuals of higher order regressions, quadratic and cubic, we can also show $\Lambda_2=0.84$ and $\Lambda_3=0.98$ where the subscripts to $\Lambda$ indicate the order of polynomial dependence for the regression fit. Obviously this is an exponential function so while $\Lambda$ would make sense in terms of the fact that the exponential function does have a linear approximation as indicated in the common Taylor expansion, $\Lambda$ does not indicate whether an explicit linear term is necessary in the overall fit for a model to explain the data. Likewise, the high $\Lambda_3$ is due to the increasing number of terms of the Taylor expansion (the first three) that are being used to approximate the exponential function using this polynomial fit. Therefore $\Lambda$ can help in understanding the dependencies with respect to various function types but is not always the best way to select the best model to fit the data.

\section{Discussion of $\Lambda$}

The introduction of $\Lambda$ helps to improve our understanding of two pressing but often unanswered questions in statistical analysis. First, what proportion of total dependence between variables is linear (or another model) and second, how can a given value of mutual information relate to the value of the correlation between variables. The second question is answered only indirectly by $\Lambda$ since direct correspondence between mutual information and correlation is still unresolved for non-normal distributions. 

$\Lambda$, it should be noted, measures for linearity of the dependence between the variables. It does not measure whether or not the joint PDF is normal distributed. In the case where $\Lambda$=1.0 and both variables have normal marginal distributions, the joint distribution is implied to be bivariate  normal. However, a value of 1.0 for variables whose marginals are not normal cannot imply anything definite about their joint distribution. In addition, a value near 1.0, though it implies linearity, does not determine the strength of the linear correlation. The correlation coefficient can be of any value, small or large, but as long as it accounts for most of the dependence, $\Lambda$ can have a high value. Likewise a large value for the correlation coefficient with a low value for $\Lambda$ indicates distinctly nonlinear behavior despite the value of the correlation coefficient.

The key value of $\Lambda$ will likely be as a secondary tool once the linearity assumption for the dependence in data is rejected. It can tell us even if the data is not linear, how much of the dependence can be explained by linear dependence and how much additional modeling is needed to explain overall dependence.

\ack{I would like to thank Dr. Cosma Shalizi for helpful comments and information regarding the normal quantile transform.}

\end{document}